Submitted to ADVANCED MATERIALS

# A Sinusoidally-Architected Helicoidal Biocomposite

N.A. Yaraghi, N. Guarín-Zapata, L.K. Grunenfelder, E. Hintsala, S. Bhowmick, J.M. Hiller, M. Betts, E.L. Principe, Jae-Young Jung, J. McKittrick, R. Wuhrer, P.D. Zavattieri, D. Kisailus

**Abstract**

A fibrous herringbone-modified helicoidal architecture is identified within the exocuticle of an impact-resistant crustacean appendage. This previously unreported composite microstructure, which features highly textured apatite mineral templated by an alpha-chitin matrix, provides enhanced stress redistribution and energy absorption over the traditional helicoidal design under compressive loading. Nanoscale toughening mechanisms are also identified using high load nanoindentation and *in-situ* TEM picoindentation.

**Graphical Abstract**

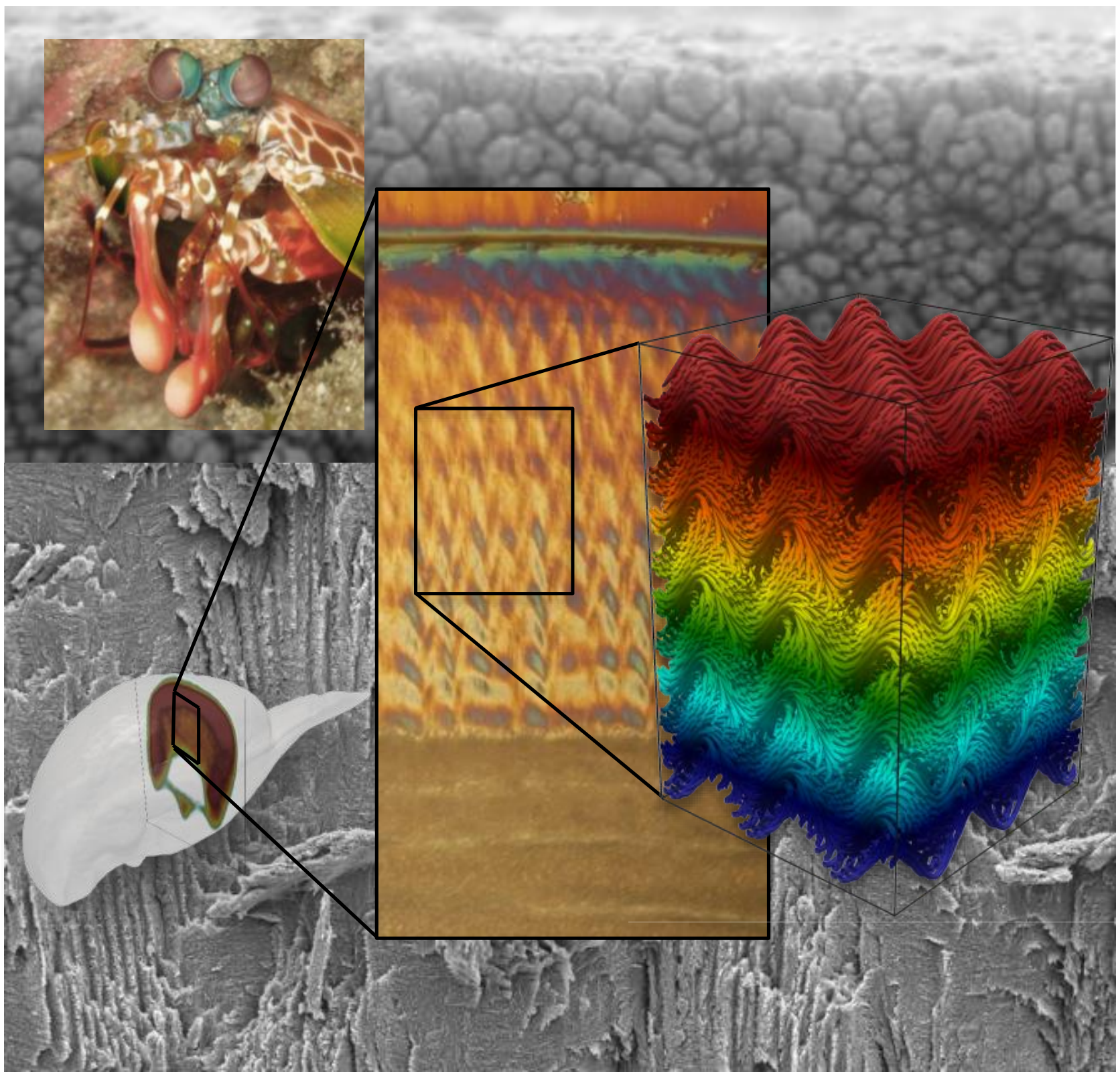

# A Sinusoidally-Architected Helicoidal Biocomposite

By *Nicholas A. Yaraghi, Nicolás Guarín-Zapata, Lessa K. Grunenfelder, Eric Hintsala, Sanjit Bhowmick, Jon M. Hiller, Mark Betts, Edward L. Principe, Jae-Young Jung, Leigh Sheppard, Richard Wuhrer, Joanna McKittrick, Pablo D. Zavattieri* and *David Kisailus**

* Prof. D. Kisailus, Nicholas A. Yaraghi
Materials Science and Engineering Program
University of California, Riverside
Riverside, CA 92521 (USA)
E-mail: david@engr.ucr.edu

Nicolás Guarín-Zapata, Prof. P.D. Zavattieri
Lyles School of Civil Engineering
Purdue University
West Lafayette, IN 47907 (USA)

Dr. L.K. Grunenfelder, Prof. D. Kisailus
Department of Chemical and Environmental Engineering
University of California
Riverside, CA 92521 (USA)

Dr. E. Hintsala
Department of Chemical Engineering and Materials Science
University of Minnesota
Minneapolis, MN 55455 (USA)

Dr. S. Bhowmick
Hysitron Inc.
Minneapolis, MN 55344 (USA)

Jon M. Hiller, Mark Betts, Dr. E.L. Principe
TESCAN USA Inc.
Pleasanton, CA 94588 (USA)

Jae-Young Jung, Prof. J. McKittrick
Materials Science and Engineering Program
University of California, San Diego
La Jolla, CA 92093 (USA)

Dr. L. Sheppard, Dr. R. Wuhrer
Advanced Materials Characterization Facility, Office of the Deputy Vice-Chancellor (R&D) Western Sydney University
Penrith, NSW 2751 (Australia)

Biologically mineralized composites offer inspiration for the design of next generation structural materials due to their low density, high strength and toughness currently unmatched by engineering technologies.[1-9] Such properties are based on the ability for the organism to utilize structural organics and acidic proteins to guide and control the mineralization process to yield hierarchical architectures with well-defined compositional gradients.

One notable example is the highly developed raptorial appendage, or dactyl, of the stomatopods, a group of aggressive marine crustaceans that use these structures for feeding upon hard-shelled and soft-bodied prey.[10-14] The dactyls of the “smashers”, those that feed primarily on hard-shelled prey, (**see Figure 1A**) takes the form of a bulbous club (Figure 1B), which is used to smash through mollusk shells, crab exoskeletons, and other tough mineralized structures with tremendous force and speed.[11-16] Achieving accelerations over 10,000*g* and reaching speeds of 23 m/s from rest, the dactyl strike is recognized as one of the fastest and most powerful impacting events observed in Nature.[11, 12] The club is capable of delivering and subsequently enduring repetitive impact forces up to 1500 N and cavitation stresses without catastrophically failing, demonstrating its utility as an exceptionally damage-tolerant natural material.

The origins of such a mechanical response lie in the structural design. Previous work identified the club as a multi-regional composite material containing an organic matrix composed of alpha-chitin fibers mineralized by amorphous forms of calcium carbonate and calcium phosphate as well as crystalline apatite.[17, 18] These investigations revealed mechanisms responsible for providing damage-tolerance and impact-resistance to the club, which were largely attributed to the interior of the club (periodic region), identified as the primary energy-absorbing layer.[17, 18] The combination of soft polymeric nanofibers and stiffer mineral provides a periodic modulus mismatch leading to crack deflection, which in combination with the helicoidal (Bouligand) architecture allows for an enhanced work of fracture due to crack twisting.[17,18] Additional work showed that this architecture provides a shear wave filtering effect during impact.[19]

The outer periphery of the club that makes direct contact with the prey (impact region) is comprised of a highly oriented and crystalline apatite mineral phase. Although it is clear that the effective transfer of impact momentum to its prey necessitates a hard and stiff outer region, details concerning the ultrastructural-mechanical property relationships of the impact region have not yet received full attention. A recent examination of crystalline and chemical aspects of the impact region suggested the presence of a fluorinated apatite-calcium sulfate phase.[20] Additionally, supporting nanomechanical studies revealed the anisotropic stiffness response and quasi-plastic nature of the impact region.[20, 21] However, key details regarding the micro- and nano-structural features and the corresponding mechanical response have yet to be revealed.

Here, we uncover a novel ultrastructural design within the impact region of the dactyl club, which exhibits a notable departure from the helicoidal structure found within most crustacean exoskeletons and affords superior mechanical advantages. We also reveal previously unreported structural and mechanical details from this region that not only provide new insights to the design of impact resistant and damage-tolerant composite materials, but also hint at the mechanisms of self-assembly and biomineralization in complex biological architectures.

Micro-computed tomographic imaging of a polished sagittal section highlights the multi-regional nature of the stomatopod dactyl club, revealing the dense and highly mineralized outer impact region and a more organic rich inner periodic region (Figure 1C). Corresponding differential interference contrast imaging (Figure 1D) of the impact region highlights a thick (~ 500 μm) bulk component and a thin (~ 70 μm) surface layer (impact surface). A close observation within this bulk section reveals a well-defined and highly ordered herringbone pattern, which takes the form of a triangular waveform (see inset Figure 1D). Each herringbone subunit has an identical wavelength, λ, equal to approximately 45 μm; however, the amplitude, A, of these units is graded in the radial direction (from 70 μm observed at the impact-periodic region interface to 100 μm at the impact region-surface interface and abruptly decreases to 50 μm within the impact surface region).

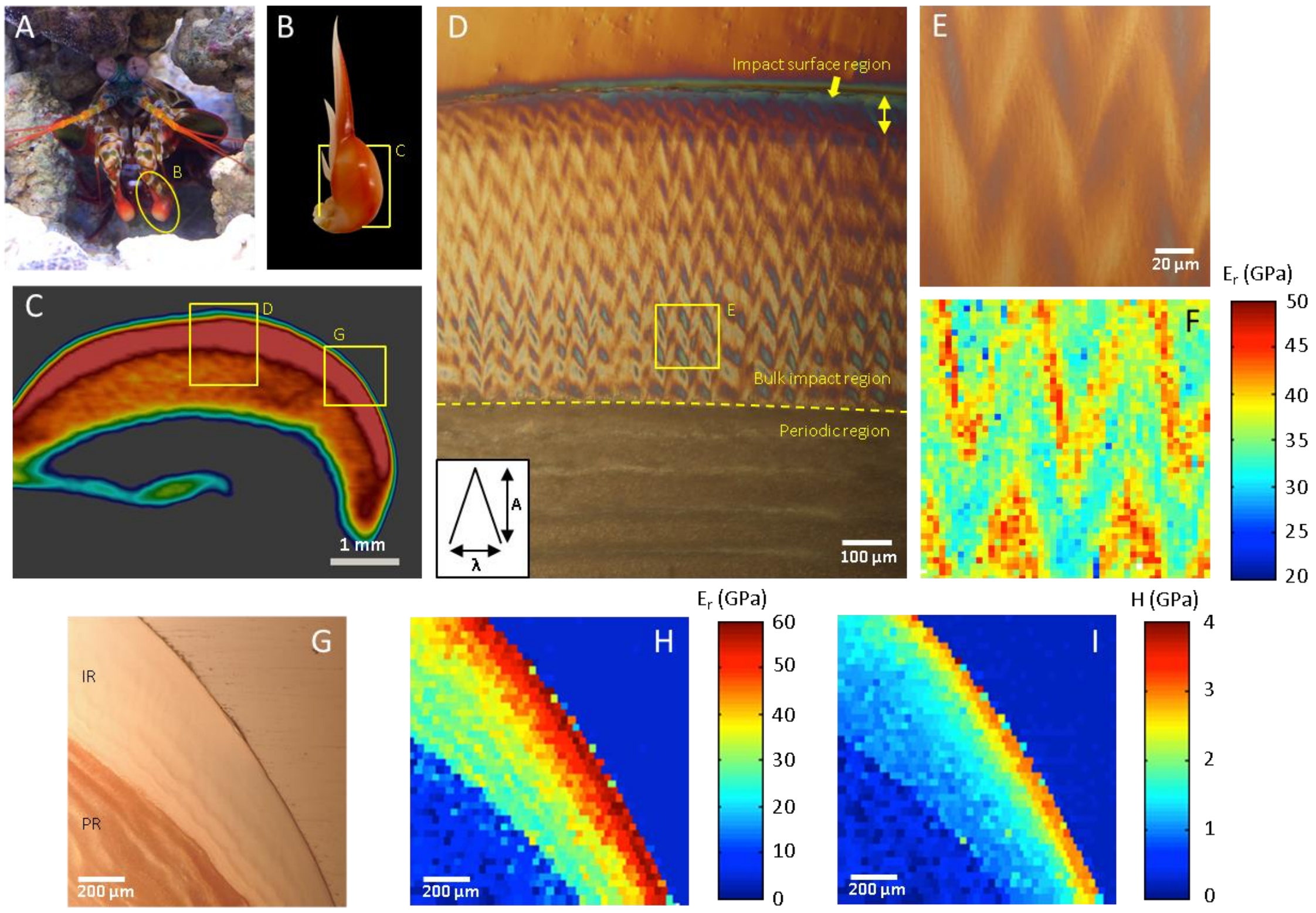

**Figure 1.** Optical microscopy and high resolution nanoindentation of the impact region. (A) Anterior of *O. scyllarus* with the dactyl segment circled in yellow. (B) Dactyl club separated from the raptorial appendage. Yellow box denotes the sagittal plane of section. (C) CT scan of a sagittal section as denoted in (B). (D) Higher magnification differential interference contrast image of region marked in (C) highlighting the impact surface, bulk impact, and periodic regions. (E) High magnification differential interference contrast image of bulk impact region as denoted in (D). (F) High resolution nanoindentation map of region (E) displaying oscillating elastic modulus correlating with herringbone pattern observed within the bulk impact region. (G) Bright field image of impact region as denoted in (C). (H,I) Low magnification nanoindentation maps of region (G) showing gradients in reduced elastic modulus (H) and hardness (I) through the impact region and near the club surface.

Nanoindentation within the impact region (Figures 1E – 1I) was subsequently used to probe local and global changes in elastic modulus and hardness within the herringbone region. High-resolution mapping (Figure 1F) reveals a strong correlation between the local reduced elastic modulus (which oscillates between approximately 30 GPa and 45 GPa) and the herringbone pattern observed from the differential interference contrast image. In addition, varied gradients in reduced modulus and hardness are observed (Figures 1H - 1I) from the periodic region into the impact region (a distance of approximately 600 microns) from 25 to 50 GPa and 0.7 to 2 GPa, respectively, and maximized at the club surface (i.e., a modulus and hardness of 60 GPa and 3 GPa, respectively). Such trends are the result of a gradient in mineralization, with calcium and phosphorus concentrations increasing towards the club surface (see Figure S1).[18] Quantitative elemental analysis via energy dispersive spectroscopy reported Ca/P molar ratios of 2.07 and 1.97 for the impact surface and impact region, respectively. This variation could be attributed to the substitution of fluorine, sulfur, and perhaps carbonate into the apatite crystal closer to the club surface, which, as reported previously, would explain the higher elastic modulus and hardness within the impact surface.[20]

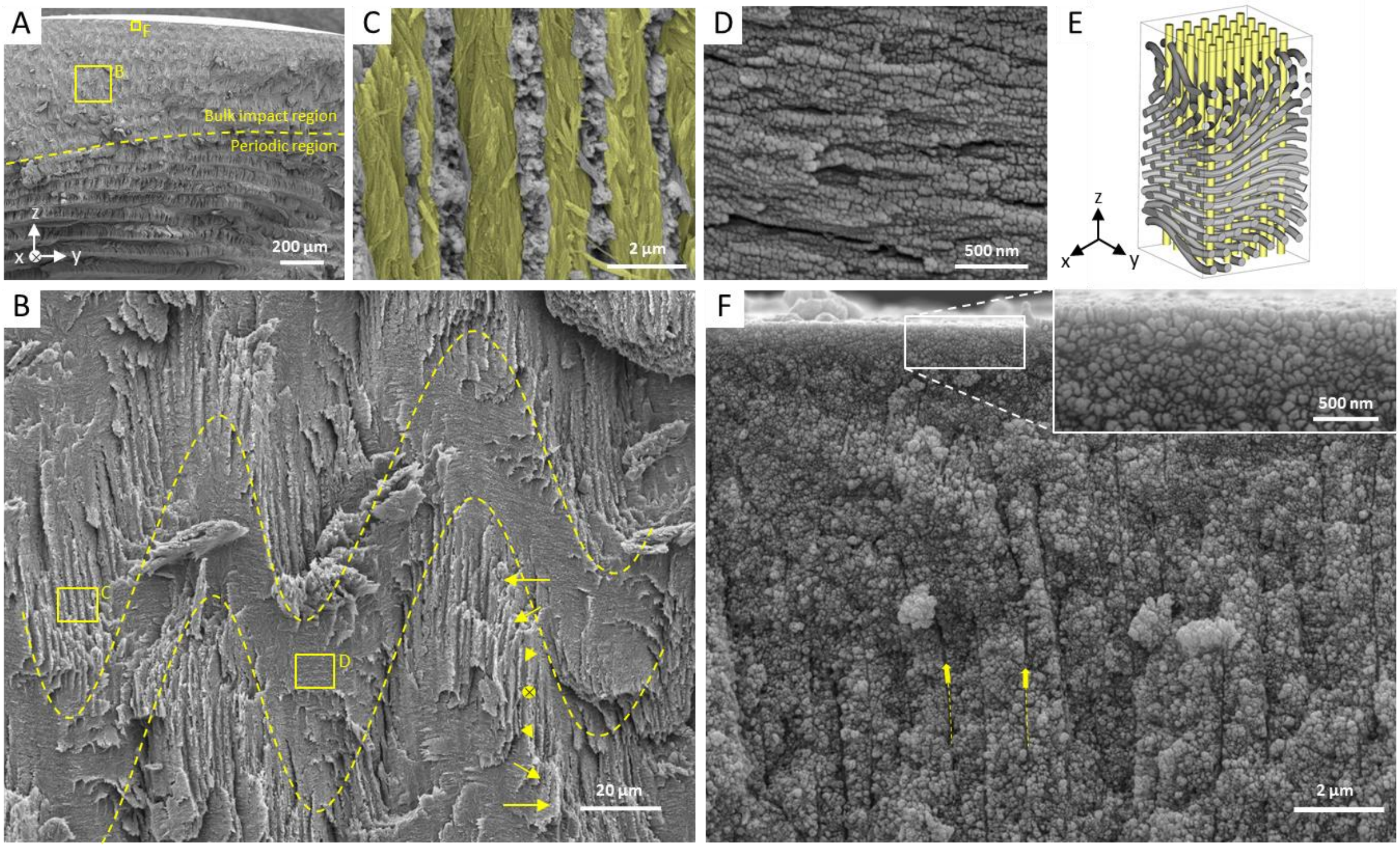

**Figure 2.** Microstructural features of the impact region. (A) SEM of fractured sagittal plane. Dashed line highlights the interface between impact and periodic regions. (B) Higher magnification of the bulk impact region as depicted in (A) showing the compacted helicoidal structure forming the herringbone motif. Dashed lines correspond to in and out of the plane fiber orientations. Arrows denote local fiber orientation. (C,D) High resolution SEM micrographs from (B) showing mineralized fibers oriented out of the plane of the page (C) and in the plane of the page (D). Out of plane oriented fibers reveal underlying network of fibrous pore canal tubules (yellow), which are aligned normal to the dactyl club surface. (E) Schematic showing the organization of rotating fibers and interpenetrating fibrous pore canal tubules. (F) Magnified area of region shown in (A) highlighting the particle morphology within the impact surface region. Arrows and dashed lines denote pore canal channels oriented normal to the club surface. Inset showing nanoparticles averaging 65 nm in diameter located at the outermost surface of the club.

Analysis of a fractured dactyl club (along its sagittal plane) by scanning electron microscopy (SEM, Figure 2) reveals that the characteristic helicoidal arrangement of mineralized chitin fibers seen in the periodic region is highly compacted laterally within the impact region, forming a herringbone pattern (Figure 2A). Fibers continue to rotate in the plane (x-y) of the micrograph (Figure 2B) about an axis (z) normal to the club surface; however, the laminations appear to be compacted in the azimuthal directions, yielding sinusoidally bent fibers forming the herringbone pattern (see also Figure S2). Closer observation (Figure 2C) highlights that the out-of-plane mineralized fibers that are oriented normal to the plane of fracture are interspersed between vertically aligned fibrous pore canal tubules, which remain in a fixed orientation normal to the surface of the dactyl club. In-plane fibers, which are oriented parallel to the plane of fracture (Figure 2D) are 49 ± 13 nm in diameter. The oriented pore canal tubules are also a common feature of the helicoidal structure, which not only function as channels for material transport during molting, but also serve an important mechanical role in terms of imparting anisotropy and toughness.[22-24] In fact, the continuation of fibrous pore canal tubules from the periodic region into the impact region likely plays a role in strengthening this interface, which can experience shear and tensile stress due to wave propagation upon impact.[19] Similar tubular structures such as those found in bone, teeth and horns have been found to enhance toughness through mechanisms such as crack deflection and resistance to microbuckling.[25, 26] A schematic depicting the intersecting rotating fiber and pore canal tubule architecture is shown in Figure 2E. Thus it is likely that areas of higher elastic modulus observed in Figure 1F correlate to the interrogation of out-of-plane fibers while lower values of elastic modulus likely result from probing in-plane fibers. This is due not only to the mechanical anisotropy of the chitinous fibers, but also due to the high degree of crystallographic texturing of the apatite mineral phase.[18, 27] The herringbone structure observed in the bulk of the impact region transitions (Figure 2F) to densely packed nanoparticles (with an average diameter of 64 ± 12 nm) with pore canal tubules persisting to the club surface.

We subsequently utilized transmission electron microscopy (TEM) to interrogate the nano-structural and crystallographic features of both the impact region and surface (Figure 3). Inspection of the impact surface (Figure 3A) confirms the observed nanoparticle morphology and array of pore canal tubules. Selected area electron diffraction (SAED, Figure 3B) and HRTEM and fast Fourier transform (FFT) analysis/ inverse fast Fourier transform (IFFT) analysis (Figure 3C, Figure S3) within the impact surface corroborates that the particles are isotropic single crystalline hydroxyapatite.

Bright-field TEM of the bulk of the impact region highlights the superimposed orthogonal fiber bundles within the herringbone structure (Figure 3D). SAED (Figures 3E and 3F) corresponding to the vertically oriented pore canal fibers and the horizontally oriented in-plane fibers, respectively, reveals the strong crystallographic texturing of the apatite mineral phase with the (002) planes showing a preferred orientation normal to the fiber direction. Thus the apatite c-axis has a preferred orientation parallel to the long axis of chitin fibers. High-resolution bright-field TEM of an isolated mineralized chitin fiber within the impact region (Figure 3G) and a corresponding Fast Fourier transform (inset, Figure 3G) reveal the lattice spacing and orientation of the (100) apatite planes, which supports the proposed texture of the mineral phase. Previous synchrotron x-ray diffraction studies proposed that the apatite mineral phase had a preferred orientation with the crystallites c-axis orientation perpendicular to the club surface; however, our TEM analyses show that this is only true for the pore canal tubule fibers.[18] In-fact, the c-axis orientation of apatite is preferentially aligned parallel to the fiber long axis and thus, for the rotating fibers, the apatite c-axis has a preferred orientation parallel to the local fiber orientation of the herringbone structure.

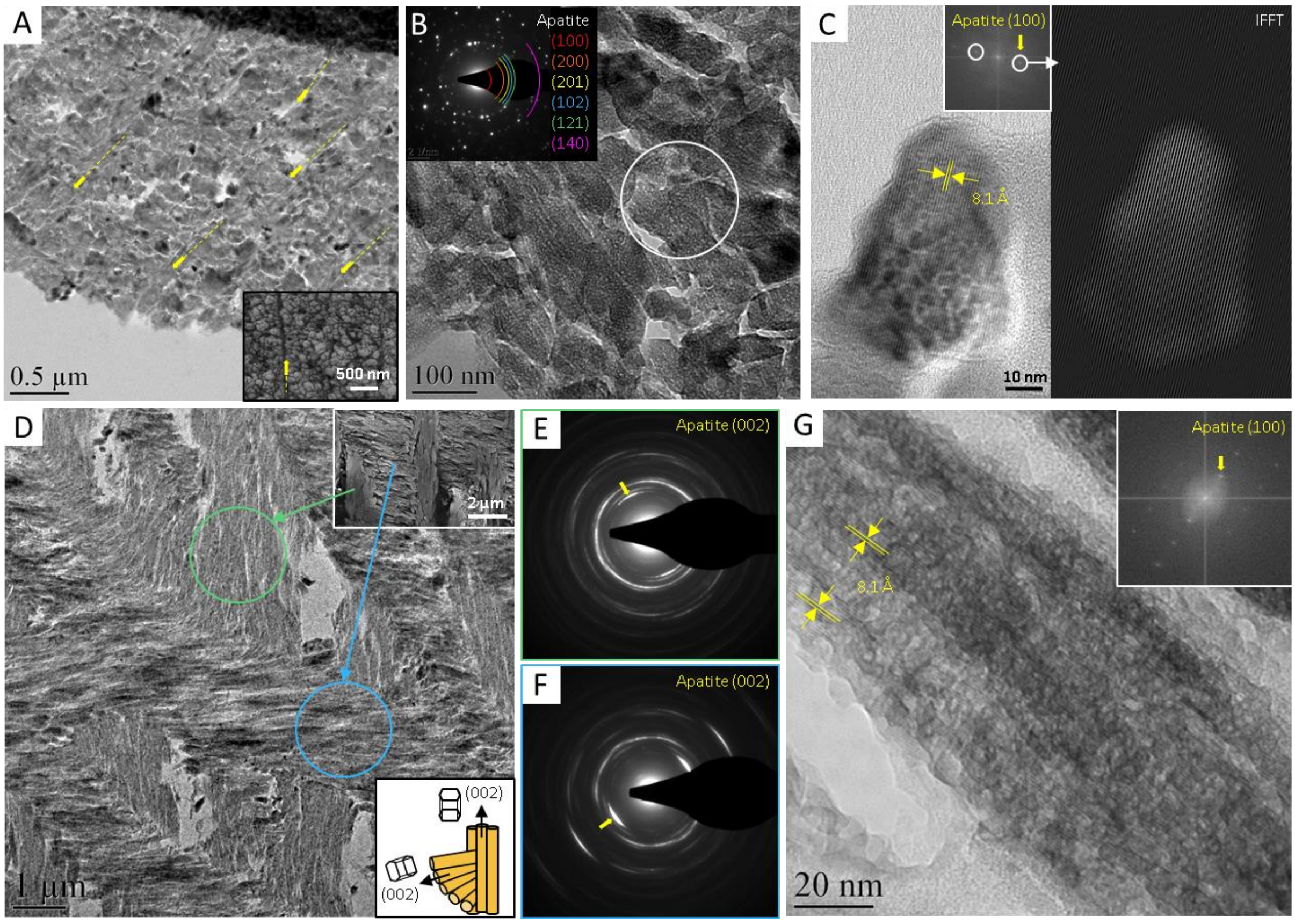

**Figure 3.** Nanostructural features of the impact region. (A) Low magnification TEM micrograph of impact surface showing pore canal fibers (marked by yellow arrows) that penetrate through to the club surface. Inset showing corresponding SEM micrograph of impact surface. (B) Higher magnification of impact surface showing nanoparticle morphology. Inset: diffraction pattern of selected area suggesting single crystalline nature of the nanoparticles. (C) (Left) High resolution TEM of an isolated nanoparticle, inset showing FFT revealing apatite (100) planes. (Right) IFFT of masked (100) planes highlighting the entire nanoparticle thus confirming single crystallinity. (D) TEM micrograph of the bulk impact region showing in-plane rotating fibers intersecting with out-of-plane pore canal fibers. Lower inset: schematic showing apatite texturing scheme and fiber architecture of the impact region, corresponding to (D). Upper inset: SEM micrograph of microtomed thin section of impact region revealing in-plane and pore canal fibers. (E,F) Selected area diffraction patterns from regions shown in (D) highlighting the preferred orientation of apatite c-axis parallel to the fiber axes for pore canal fibers (E) and in-plane rotating fibers (F). (G) Higher magnification of single mineralized fiber from (D). Inset: Fast Fourier Transform of (G) highlighting apatite (100) planes oriented parallel to the fiber long axis.

High load nanoindentation within the bulk of the impact region was used to initiate fracture and examine the effect of local microstructure on crack propagation in order to identify potential toughening mechanisms of the herringbone structure. A representative SEM micrograph (Figure 4A) of a 1000 mN peak load indent placed within the bulk of the impact region, which was subsequently etched in dilute acid to highlight structural features, reveals contrast due to the preferential etching of out-of-plane fibers of the herringbone structure. Cracks are observed emanating from the area of the indent; however, these cracks appear to become arrested at the transitional zones between in-plane and out-of-plane oriented fibers. Examination further away from the indent shows that cracks re-appear within areas containing out-of-plane fibers. In fact, cracking is observed only in regions where the local fiber orientation is normal to the plane of the section. It is expected that cracks twist as well, following the local orientation of helicoidally arranged fibers. Similar observations have been made with respect to the periodic region of the dactyl club and helicoidal structures found within most arthropod cuticles as well as other natural composite materials including lamellar bone, cell walls of wood and fish scales.[17, 18, 28-33] Crack deflection is an extrinsic form of toughening that is well-documented in natural composite materials, specifically biomineralized tissues.[3] The periodic nature of hard and soft interfaces, in this case between alpha-chitin fibrils and hydroxyapatite crystals, results in a crack-tip shielding effect that changes the crack driving force and thereby arresting crack propagation.[35, 36] He and Hutchinson discussed the role of elastic mismatch on the strain energy release rate of a crack, which determines if the crack gets deflected at an interface or penetrates through to the other solid.[34]

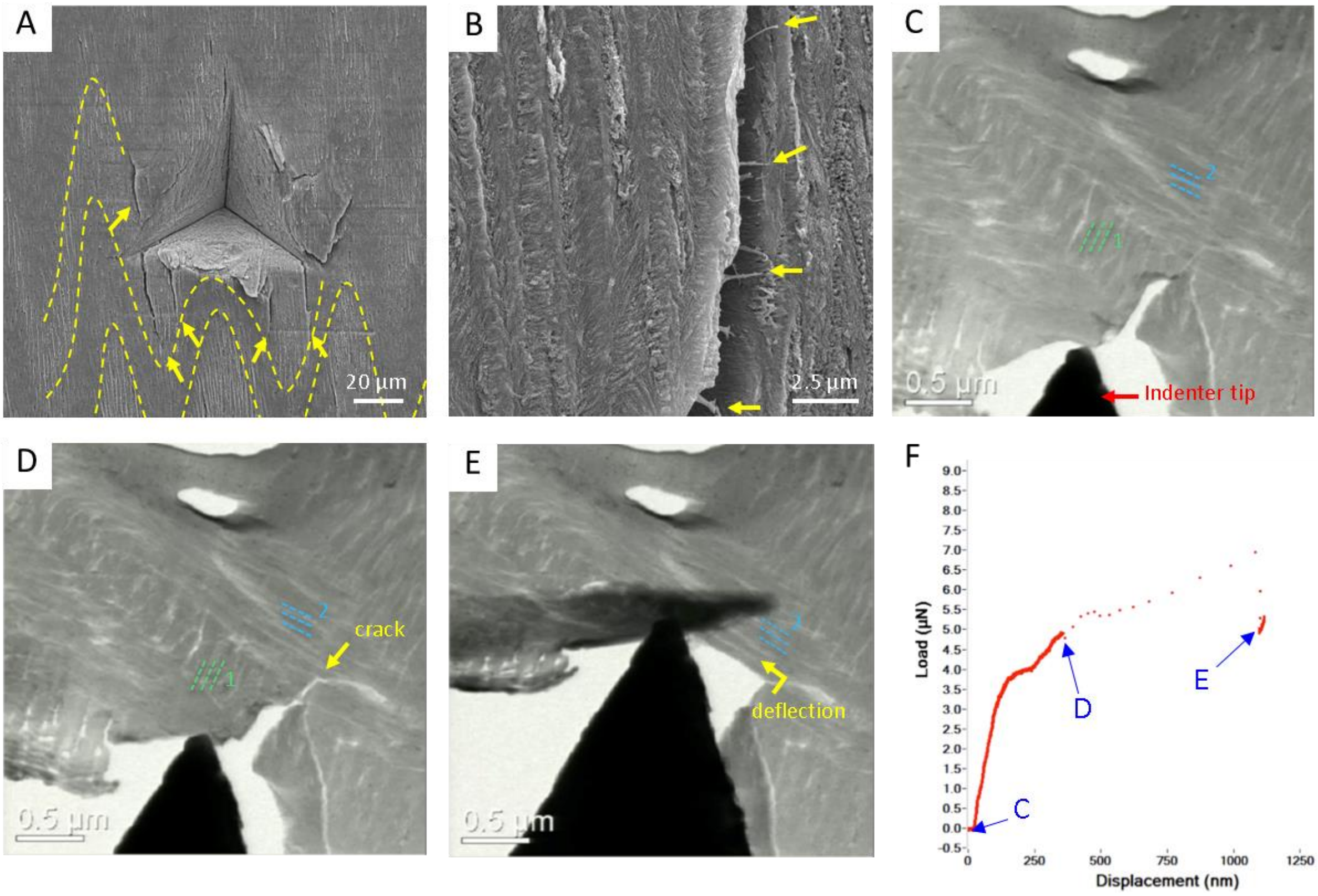

**Figure 4.** High load nanoindentation and *in-situ* TEM picoindentation of the bulk impact region. (A) SEM micrograph of 1000 mN peak load indent placed within the impact region. Surface was lightly etched (see methods) to reveal the herringbone structure and crack deflection (arrows) at interfaces between in-plane and out-of-plane fiber orientations (dashed line). (B) SEM micrograph showing fiber bridging (yellow arrows) at the indent edge. (C-E) TEM micrographs showing progressing stages of loading of a FIB-sectioned area of the bulk impact region. The local microstructure consists of overlapping fiber bundles that are oriented normal to one another. Dashed blue and green lines depict local fiber orientation from two separate overlapping bundles (1 and 2, respectively), which correspond to rotating and pore canal fibers of the herringbone structure. (F) Load-displacement curve at various stages of indentation corresponding to (C-E). (C,D) Mode I crack is opened at pre-existing notch in region 1 and propagates in the direction of the local fibers (dashed green lines). (D) Crack approaches fiber bundle 2 and begins to deflect at an off-angle. (E) Localized failure as crack transitions to mode III failure due to out-of-plane bending. Crack is deflected 90° in the direction of bundle 2 fibers (dashed blue lines).

However, unlike a simple 90° crack deflection, cracks within this structure are twisting and thus increase toughening. We hypothesize that, as opposed to crack deflection in interfaces as observed in other biological materials like nacre, crack twisting in a helicoidal structure can provide (i) a strategy to create multiple twisting microcracks that grow at different nucleation sites but never coalesce given the nature of the helicoidal architecture and (ii) more crack surface per unit volume, and therefore maximizing energy dissipation.[17,18] In addition to crack deflection/rotation, we also observe fibrils bridging an opened crack surface adjacent to the indented area (Figure 4B). These fibers act to reduce the crack-tip stress concentration and prevent further crack opening, providing an additional source of extrinsic toughening.[4, 37, 38] Finally, we expect that the herringbone structure offers an additional improvement in toughening over a helicoidal motif by incorporating a sinusoidal interface between lamellar fibrillar layers.

In order to examine *in-situ* fracture and identify toughening mechanisms at the nanoscale within the bulk impact region, TEM pico-indentation was performed on a focused-ion beam (FIB) milled section. Figures 4C - 4F show the results of progressing stages of quasi-static loading of a thin section containing overlapping fiber bundles (correspond to the in-plane and pore canal tubule fibers, Figure 4C) that are oriented normal to one another. Initially, a mode-I crack is opened at a pre-existing notch located within fiber bundle 1 (green dashed lines, corresponding to the in-plane mineralized fibers) and continues to propagate parallel to the long axis of these fibers (Figure 4D). After reaching a load of approximately 5 μN, the crack front approaches fiber bundle 2, which corresponds to a fibrous pore canal tubule (blue dashed lines) oriented normal to the direction of crack propagation. Beyond this load, the material fails locally as the crack is deflected and transitions to mode-III failure due to out-of-plane bending (Figure 4E). Direct observations from the TEM micrograph show that the crack was deflected 90° in the direction of the pore canal fiber bundle. Load-displacement data of the indentation event (Figure 4F) highlights an initial linear-elastic region followed by a pop-in event, which corresponds to the crack opening as shown in Figure 4D. These results not only provide direct evidence of nanoscale extrinsic toughening through crack deflection at an orthogonal fiber interface, but also highlight the effect of crystallographic texturing (results from Figure 3) on imparting mechanical anisotropy to polymeric fibers.

Finite element (FE) analysis in combination with 3D printing was used to examine the role of the herringbone architecture on the mechanical performance under compression versus a helicoidal structure. Schematics of the helicoid and a herringbone architecture are shown in Figure 5A (left and right, respectively). The coloration denotes the relative position of the fibers along the z-axis. The herringbone microstructure was modeled using a transverse isotropic material, changing the orientation of the symmetry axis for each element.

This alignment follows the tangent of the curve formed by the intersection of Equation (1)

$$f(x, y) = A\sin\left(\frac{2\pi x}{\lambda}\right)\sin\left(\frac{2\pi y}{\lambda}\right) \qquad (1)$$

and a vertical plane rotated by the pitch angle. Both helicoidal and herringbone structures were subjected to uniaxial strain conditions in the *z* direction. Figures 5C and 5D show surfaces of constant normalized von Mises stress ($\sigma_{\mathrm{Mises}}/E_t$) for $1.89\times10^{-2}$, $1.95\times10^{-2}$, and $2.28\times10^{-2}$ subjected to compressive loading for the Bouligand and herringbone structures, respectively. The von Mises stress is selected as figure of merit since it is an invariant of the stress tensor that accounts for the deviatoric components, i.e., for energy spent in shear deformations.[50] Since the impact region is under compression, shear stresses would be of major relevance for the composite than volumetric stresses. These results reveal that the herringbone structure allows for a greater redistribution of stresses compared to the helicoidal design. Our hypothesis is that the redistribution of stresses is directly translated into a redistribution of damage, opposed to a localization, which would lead to catastrophic failure.

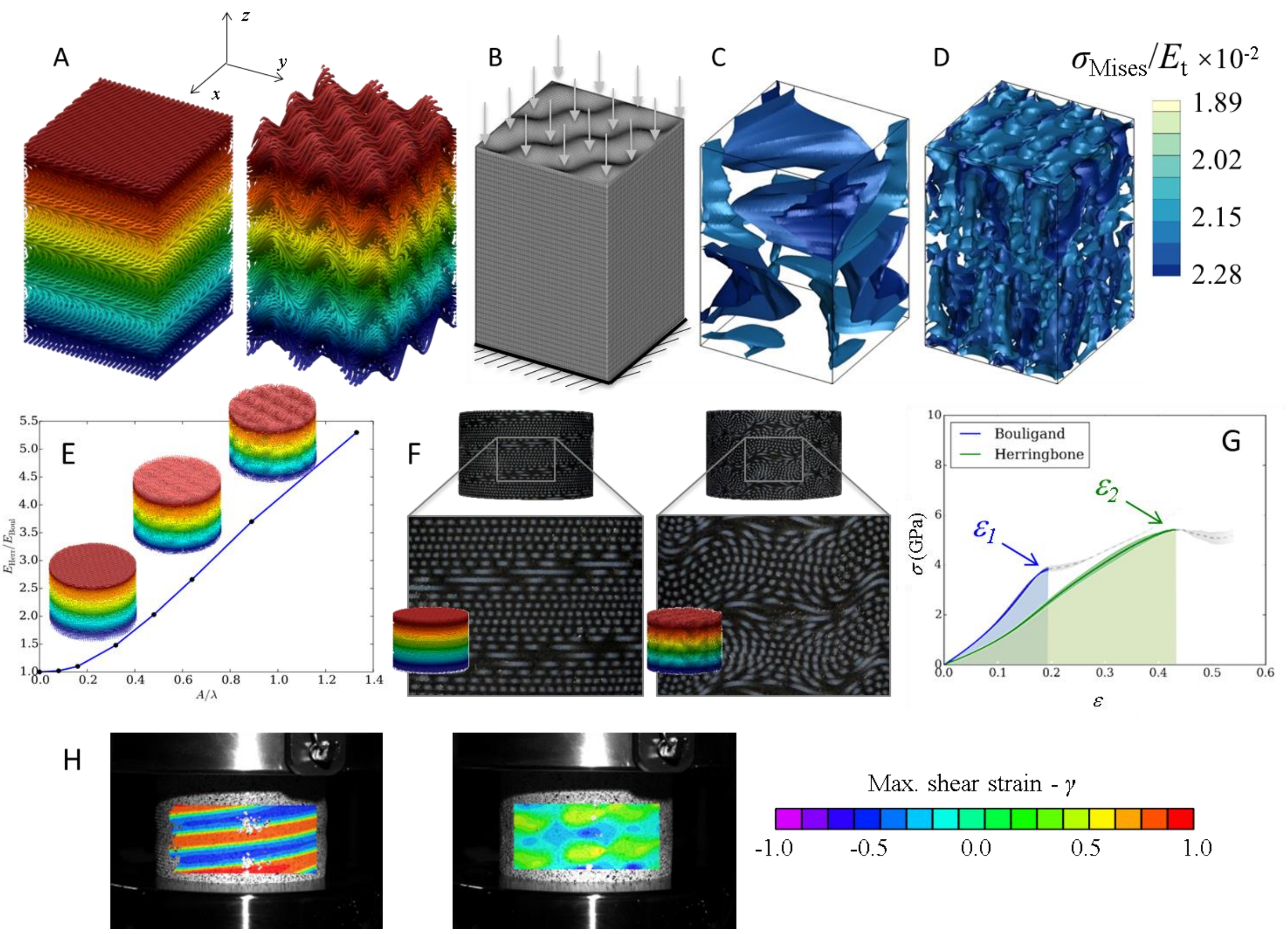

**Figure 5.** Finite Element Analysis and testing of 3D printed mimics comparing the helicoidal and herringbone structures. (A) Schematic of the geometry and fiber orientation for the helicoidal and herringbone structures. The colors represent the relative position along the *z*-axis. (B) A mesh for the case of the herringbone structure used for the analysis. (C-D) Surfaces of constant normalized von Mises stress ($\sigma_{Mises}/E_t$) for the values 1.89×$10^{-2}$, 1.95×$10^{-2}$, and 2.28×$10^{-2}$ for the helicodial (C) and herringbone (D) cases. Observe the von Mises stresses that are redistributed within the volume. (E) Relative Young modulus for herringbone pattern with respect to the helicoidal case. The horizontal axis shows the aspect ratio between the amplitude and wavelength of the herringbone pattern. (F) 3D printed samples of the helicoid and herringbone structures. (G) Results of compression tests for the 3D printed samples. (H) Comparison for the 3D printed samples at a deformation of 0.1 (left, helicoidal; right, herringbone).

The effect of the amplitude to wavelength ratio, $A/\lambda$, of the herringbone structure on the mechanical properties was also examined under uniaxial stress loading conditions. We recall that the herringbone structure within the impact region of the dactyl club exhibits increasing amplitude moving from the impact-periodic interface to the club surface. This is evident by differential interference contrast imaging at this interface (Figure S7) showing flat layers within the periodic region gradually becoming wavy, then finally forming a well-defined herringbone pattern within the impact region. We hypothesize that certain crystallization stresses occur during the transformation from a hydrated amorphous state (within the periodic region) to hydroxylapatite (within the impact region) driving the formation of the herringbone-like structure. Figure 5E shows that the normalized effective Young's modulus of the herringbone structure increases with $A/\lambda$ (note that the helicoid corresponds to $A/\lambda$ = 0). While nanoindentation in the sagittal plane (Figure 1H) shows a similar trend of increasing $E_r$ moving towards the club surface, it is likely that this is due to the gradient in mineralization (increasing mineral content near the club surface, see Supplemental Figure S1). Nonetheless, we hypothesize that the amplitude gradient of the herringbone structure also contributes to a gradient in Young's modulus in the loading direction of the club. This structural gradient would ensure that there is no abrupt interfacial plane where the material properties completely and drastically change from one region to the other, which likely plays an important role in preventing cracks from initiating at this interface thus causing the material to fail catastrophically. Previous studies have shown that increasing the amplitude to wavelength ($A/\lambda$) ratio of sinusoidal interfaces in materials can delay unstable crack propagation and increase toughness.[39, 40] Our results may provide new insights on how to impart additional toughness as well as modulate through-thickness stiffness in composite laminates by varying the architecture of in-plane fibers. Carbon-fiber epoxy composite laminates mimicking the helicoidal architecture of the periodic region have been found to exhibit improved impact-resistance in comparison to traditional quasi-isotropic aerospace designs.[41]

3D printed biomimetic herringbone and helicoid structures were fabricated and tested to failure under compressive loading. Figure 5F shows the 3D printed helicoid (left) and herringbone (right) cylindrical parts. For both structures, the fibers are printed with a hard polymer and the matrix with a soft elastomer (see experimental section). The samples were subsequently coated with paint to obtain a random black-and-white speckle pattern for posteriori 3D digital image correlation (3D-DIC) analysis. The experimental load vs. displacement experiments allow large deformations in the matrix and fibers (Figure 5G), which ultimately leads to a relatively more compliant herringbone structure as compared with the helicoidal specimen. This is due to the fact that the sinusoidally-arranged fibers within the 3D printed mimics straighten upon compression. Additionally, experiments show that the herringbone structure has better energy absorption capabilities than the helicoid motif. Both curves show a local maximum in stress at $\varepsilon_1$ and $\varepsilon_2$ for the helicoid and herringbone structures, respectively. This local maximum corresponds to critical applied strain where significant damage is observed. The fact that $\varepsilon_1 < \varepsilon_2$ suggests that the herringbone structure is capable of withstanding large strains without significant damage leading to a higher absorbed energy density. The values are 0.35 $J{\cdot}m^{-3}$ and 1.20 $J{\cdot}m^{-3}$, for the helicoid and herringbone structures, an increase of 3.43 times. In fact, Figure S8 shows more damage in the helicoidal structure. 3D DIC of each sample at an applied strain of 0.1 (Figure 5H) shows more strain localization in the helicoidal structure. On the other hand, the herringbone structure exhibits much lower and distributed strains, which also enables the material to withstand localized damage and efficiently transfer momentum to the surface it is striking.

High load nanoindentation in combination with FE simulation was also used to ascertain the role of the nanoparticulated impact surface on the mechanical response of dactyl club exocuticle. Figure 6A shows a dark-field optical micrograph of a polished sagittal section of the dactyl exocuticle with a corresponding schematic, highlighting the nanoparticle-based impact surface interfacing with the herringbone structure within the impact region. High load indentation was conducted within the impact surface and the impact region using an embedded dactyl club polished in the coronal direction through the impact region. SEM observations (Figure 6B) of quasi-static indentations to peak loads of 100 mN, 500 mN, and 1000 mN within the impact surface (top row) and impact region (bottom row) show clear differences in the deformation and fracture behavior. Indents within the impact surface layer generally feature a high degree of pile-up surrounding the faces of the indent. This pile-up behavior is not observed within the impact region (i.e., with the impact surface removed); in general, we see smooth faced plastic-looking indents with minimal cracking within the impact region. We note that the cracks observed in the bottom row of Figure 6B were the result of inadvertent stresses introduced during polishing and were not, in fact, the result of indentation. However, cracking is observed within the impact surface for all three indents. Similar pile-up behavior has been attributed to shear-band localization and that the impact surface consisted of an oriented fluorinated apatite rod-like microstructure through which the shear induced cracking led to the rotation and sliding of crystallites.[20, 21] However, our HRTEM analyses reveal that the particles within the impact surface are single crystalline and isotropic domains that are crystallographically separated. Given that the pile-up behavior is not apparent within the herringbone-structured impact region, we suspect that the pile-up behavior in response to indentation is the result of meso-scale displacement of the isotropic nanoparticles. This motion may have implications for the redistribution of stress during impact loading. The ability for nanoparticles to plastically deform through particle motion may be critical for preventing stress concentration during impact loading. Without the impact surface, peaks of the underlying herringbone structure would likely serve as sites for stress concentration and potentially crack initiation. Nanoparticulated surface coatings have been observed in other biological composite materials such as the ultra-hard teeth of the chiton, *Cryptochiton stelleri*, a marine mollusk.[42] These teeth feature an outermost 2 μm thick layer of densely packed sub-100 nm magnetite particles with no preferred orientation, which cover an underlying rod-like microstructure and provide abrasion resistance.[42] The incorporation of ceramic nanoparticles into epoxy polymers used as matrix materials for fiber-reinforced composites has also been shown to improve the toughness as well as Young's modulus of the polymer.[43-45] The extent of plastic deformation of the epoxy is enhanced so as to dissipate more energy in regions surrounding the crack tip.[43, 44]

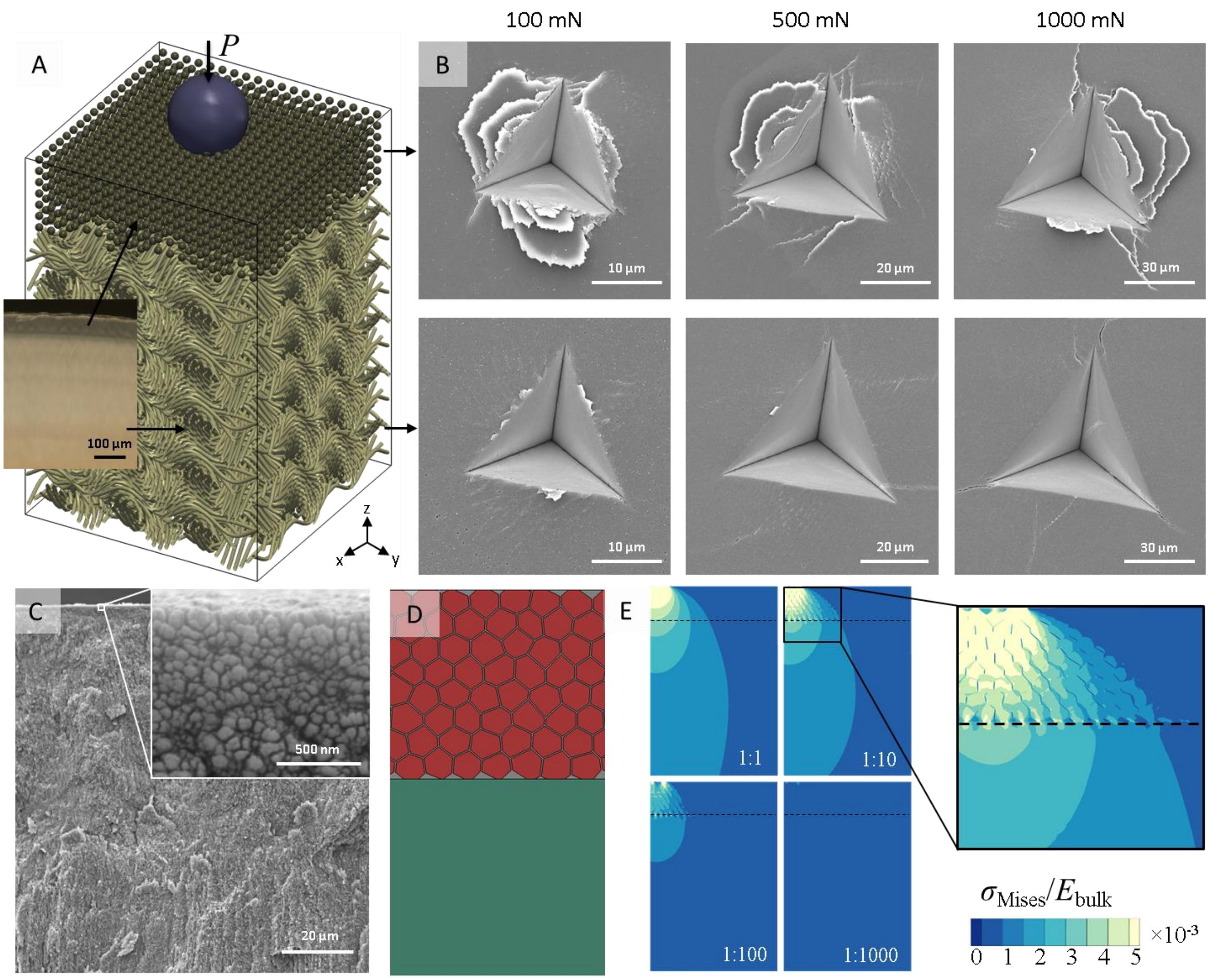

**Figure 6.** High load nanoindentation and FE Simulation of the impact surface particulate layer. (A) Three dimensional schematic of the impact region and impact surface showing the particulate layer

placed on top of the herringbone structure. Inset: dark field optical micrograph of the corresponding area. (B) SEM micrographs of high-load indents placed within the impact surface and the impact region to peak loads of 100 mN, 500 mN, and 1000 mN. (C) SEM micrographs of the impact surface highlighting the nanoparticle morphology. (D) Schematic model for the particulated layer used in the FE Simulation. (E) von Mises Stress in the region highlighted in (D). The particulated region is delimited by the dashed line. The Young modulus ratio between matrix and particle is changed: 1/1, 1/10, 1/100, 1/1000. We can see a confinement of the stress when the ratio between the particles and matrix increases.

FE simulations were used to examine the effect of the nanoparticles of the impact surface (Figure 6D,E) on the mechanical response of the dactyl club exocuticle. A SEM micrograph of the impact surface and schematic model used for FE simulation are shown in Figures 6C and 6D, respectively. The model consists of a 70 µm thick layer of isotropic particles with 88% volume fraction. This layer is placed on top of an isotropic homogeneous substrate. The material properties of the upper layer are chosen to match those of the substrate, using the Voigt homogenization method.[46] The simulation was displacement controlled, and the images in Figure 6E are shown at 1 µm; the size of the loading region was 10 µm. Figure 6E shows the distribution of von Mises stress for cases where the Young's modulus ratios between the particle and matrix were varied from 1:1, 1:10, 1:100, and 1:1000. The 1:1 case signifies no particles within the outer layer. In the stomatopod, we expect the ratio to be in the range of 1:100 to 1:1000, representing particles that are stiffer than the material between them. We can see that stress is confined within the impact surface layer when the ratio between the Young's modulus of the particles and matrix material is increased, which supports our preliminary hypothesis concerning the role of particles in redistributing stress.

## Conclusions

The extraordinary designs in the tough and energy absorbent core and stiff and hard exterior in the stomatopod dactyl club yield exceptional damage tolerance from repetitive high-energy impact events. Here, we uncovered a previously unreported ultrastructural motif within the impact region, featuring a highly ordered compacted and pitch-graded sinusoidal arrangement of helicoidally arranged alpha-chitin fibers mineralized with highly textured apatite mineral. TEM analysis showed that the apatite crystallites exhibit a preferred orientation parallel to the long axis of the chitin fibers, which imparts mechanical anisotropy. FE analysis and testing of biomimetic 3D printed structures revealed that the herringbone architecture offers an enhancement in stress redistribution and out-of-plane stiffness in response to compressive loading compared to the helicoidal architecture. Although the FE models did not take into account the pore canal tubule network, we expect that the out-of-plane mineralized fibers in combination with the sinusoidal arrangement of helicoidal in-plane mineralized fibers provides an optimal out-of-plane compressive strength and toughness necessary for transferring maximum load to its prey while resisting fracture. Additionally, the incorporation of sinusoidal interfaces in the herringbone structure may improve toughness via extending the path length for crack growth, thereby enhancing energy dissipation. We also identified a thin layer of isotropic single crystalline apatite nanoparticles, which cap the impact region and likely provide additional stress redistribution through particle motion. Questions still remain as to the mechanisms driving the formation of the herringbone structure and the potential role amorphous and crystalline mineral phases may play in driving the assembly of helicoid versus herringbone structures. Future work will investigate the crystallization and formation of the herringbone structure. In addition, we are now demonstrating the ability to mimic these complex structures through 3D printing and beyond. Nonetheless, the findings presented herein can provide inspiration and ongoing design guidelines for the fabrication of next-generation impact-resistant composite materials.

**Experimental Section**

*Research Specimens*: Live specimens of *Odonodactylus scyllarus* were obtained from a commercial supplier and housed in an artificial seawater system. Fresh inter-molt dactyl clubs were obtained from both live and recently deceased specimens. Polished cross-sections of the dactyl club were prepared by embedding in epoxy (System 2000, Fibreglast, USA), sectioned using a low speed saw with diamond blade, and polished with progressively finer SiC and diamond abrasive down to 50 nm grit. Fractured samples were obtained along the sagittal plane using a hammer and sharpened chisel. To highlight interfaces of the fibrous microstructure in the herringbone structure from polished sections, some specimens were washed in aqueous 5% acetic acid solution for 30 seconds to partially demineralize the sample surface. Samples were then washed in DI water and dried in air prior to SEM observation.

*Nanoindentation:* Nanoindentation on flat polished cross-sections of the dactyl club was performed at room temperature using a TI 950 TriboIndenter (Hysitron, USA). Indentation mapping of the impact region was performed using a low-load transducer with a diamond cube corner tip. Indents were placed in a square array and controlled in displacement to a depth of 200 nm. The trapezoidal load function consisted of a 5 second load, followed by a 2 second hold, and a 5 second unload. High-resolution mapping was performed on a 90 μm by 90 μm area and indents were spaced 3 μm apart. Lower resolution mapping was performed on a 1.2 mm by 1.2 mm area and indents were spaced 30 μm apart. Values for reduced elastic modulus and hardness were calculated using the Oliver and Pharr method [47]. Spatial maps of reduced modulus and hardness were subsequently plotted using the scatter function in MATLAB (MathWorks, USA). High load indents were performed using a high load transducer with a diamond cube corner tip to induce fracture. Indents were placed individually in areas of the impact region and impact surface and were controlled in load. The load function consisted of a 5 second load followed by a 5 second unload.

*Scanning Electron Microscopy:* Fractured, microtomed, and polished indented sections of the dactyl club were mounted to aluminum pin mounts using carbon tape and sputter coated with a thin layer of platinum and palladium. In some cases, conductive silver paint was also applied to increase conductivity and prevent charging. Specimens were imaged at 10 kV using an XL-30 FEG (FEI-Philips, USA).

*X-Ray Mapping:* Polished sections of the club were mounted on carbon tape-coated aluminum stubs, carbon coated, and analyzed using a JSM-840 SEM (JEOL, USA) operating at 20 kV. X-ray maps were post-processed using the "Chemical Imaging" software package within the Moran Scientific Microanalysis System [48].

*Transmission Electron Microscopy:* Specimens for TEM were prepared by first isolating 1 mm cubed pieces of the impact region, which were cut from a transverse section of the dactyl club using a razor blade. The pieces were then fixed with glutaraldehyde (2.5%) in aqueous sodium phosphate buffer solution (0.1M, pH = 7.2) for 2 hours and subsequently washed in DI water three times for 5 minutes each. Specimens were then post-fixed in osmium tetroxide (1%) in sodium phosphate buffer (0.1M, pH = 7.2) overnight and washed again in DI water three times for 5 minutes each. Samples were serially dehydrated to 100% ethanol and embedded in resin (Epofix Cold-Setting Embedding Resin, Electron Microscopy Sciences, USA) in silicon molds at room temperature overnight. Cured resin blocks were then sectioned using an ultramicrotome (RMC MT-X, Boeckeler Instruments, USA) and diamond knife (PELCO, Ted Pella, USA) to produce 70 nm thin sections, which were deposited on carbon-coated copper grids and imaged at 300 kV in a CM300 TEM (FEI-Philips, USA).

*In-situ TEM Picoindentation:* Thin sections of the bulk impact region were milled from a polished transverse section using a LYRA3 FIB-SEM (TESCAN, USA). *In-situ* indentation was performed using a PI-95 TEM Picoindenter (Hysitron, USA) and imaged in a Tecnai F30 TEM (FEI, USA) at 300 kV. Samples were indented using a boron-doped diamond wedge tip with 100 nm radius of curvature and controlled under load at a rate of 0.18 μN/s.

*3D Printing & Bulk Mechanical Testing:* Cylindrical samples were printed in a Connex350™ (Objet, USA), using VeroWhitePlus for the fibers and TangoBlackPlus for the matrix. In both cases the cylinders are 60 mm in diameter and 37 mm in height. The volume fractions are 0.242 and 0.267, for the Bouligand and herringbone, respectively. The samples were tested under compressive load in a MTS Insight 300 (MTS System Corporation, USA) using a 569331-01 load cell, the rate used was 0.5 mm/min. The software VIC3D from Correlated Solutions was employed for the 3D Digital Image Correlation analysis.

*3D Finite Element Modeling (FEM):* The FEM simulations were carried out in Abaqus/Standard [49]. For the herringbone structure, we used 81×81×81 trilinear elements (C3D8), and the body was subjected to uniaxial strain conditions in the *z* direction. In the case of the particulated model, we used 594602 triangular linear elements (CPS3).

**Acknowledgements**

D.K. would like to acknowledge financial support from the Air Force Office of Scientific Research (AFOSR-FA9550-12-1-0245) and the Air Force Office of Scientific Research, Multi-University Research Initiative (AFOSR-FA9550-15-1-0009). D.K. would also like to acknowledge financial support from the AFOSR DURIP Grant FA9550-10-1-0322 for the nanoindenter. This research was conducted with Government support under and awarded by DoD, Air Force Office of Scientific Research, National Defense Science and Engineering Graduate (NDSEG) Fellowship, 32 CFR 168a. P.Z. would like to acknowledge financial support from National Science Foundation CAREER award CMMI 1254864 and the AFOSR DURIP Grant FA2386-12-1-3020 for the 3D printer. Electron microscopy was performed in the Central Facility for Advanced Microscopy and Microanaylsis at UC Riverside. The authors gratefully acknowledge the technical assistance of Mathias Rommelfanger and Dr. Krassimir N. Bozhilov from UC Riverside, Drew Erwin from TESCAN USA Inc., Joseph Lefebvre, Jeff Rollings, and Ryan J. Stromberg from Hysitron Inc., and Chanhue Jeong, and David Restrepo from Purdue University. The authors would also like to acknowledge Dr. Garrett Milliron for his initial observations and contributions to this work. Christopher Salinas, Steven Herrera, and Jesus Rivera from the Kisailus group are acknowledged for helpful discussions.

**References**

[1] C. Sanchez, H. Arribart, M.-M. Giraud-Guille, *Nat. Mater.* **2005**, 4, 277.

[2] H.-B. Yao, H.-Y. Fang, X.-H. Wang, S.-H. Yu, *Chemical Society Reviews* **2011**, 40, 3764.

[3] R. O. Ritchie, *Nat. Mater.* **2011**, 10, 817.

[4] U. G. Wegst, H. Bai, E. Saiz, A. P. Tomsia, R. O. Ritchie, *Nat. Mater.* **2015**, 14, 23.

[5] S. E. Naleway, M. M. Porter, J. McKittrick, M. A. Meyers, *Adv. Mater.* **2015**, 27, 5455.

[6] P. Fratzl, R. Weinkamer, *Progress in Materials Science* **2007**, 52, 1263.

[7] P.-Y. Chen, J. McKittrick, M. A. Meyers, *Progress in Materials Science* **2012**, 57, 1492.

[8] M. A. Meyers, P.-Y. Chen, A. Y.-M. Lin, Y. Seki, *Progress in Materials Science* **2008**, 53, 1.

[9] M. A. Meyers, J. McKittrick, P.-Y. Chen, *Science* **2013**, 339, 773.

[10] R. Caldwell, H. Dingle, *Naturwissenschaften* **1975**, 62, 214.

[11] S. N. Patek, W. L. Korff, R. L. Caldwell, *Nature* **2004**, 428, 819.

[12] S. N. Patek, R. L. Caldwell, *J. Exp. Biol.* **2005**, 208, 3655.

[13] J. R. Taylor, S. N. Patek, *J Exp Biol* 2010, 213, 3496.

[14] E. A. K. Murphy, S. N. Patek, *The Journal of Experimental Biology* **2012**, 213, 4374.

[15] S.N. Patek, M. V. Rosario, J. R. A. Taylor, *Journal of Experimental Biology* **2013**, 216, 1317.

[16] M. Tadayon, S. Amini, A. Masic, A. Miserez, *Advanced Functional Materials* **2015**, 25, 6437.

[17] G. Milliron, "Lightweight Impact-Resistant Composite Materials: Lessons from Mantis Shrimp." **2012.**

[18] J. C. Weaver, G. W. Milliron, A. Miserez, K. Evans-Lutterodt, S. Herrera, I. Gallana, W. J. Mershon, B. Swanson, P. Zavattieri, E. DiMasi, D. Kisailus, *Science* **2012**, 336, 1275.

[19] N. Guarín-Zapata, J. Gomez, N. Yaraghi, D. Kisailus, P. D. Zavattieri, *Acta Biomaterialia* **2015**, 23, 11.

[20] S. Amini, A. Masic, L. Bertinetti, J. S. Teguh, J. S. Herrin, X. Zhu, H. Su, A. Miserez, *Nat. Commun*. **2014**, 5, 3187.

[21] S. Amini, M. Tadayon, S. Idapalapati, A. Miserez, *Nat. Mater.* **2015**, 14, 943.

[22] C. Sachs, H. Fabritius, D. Raabe, *J. Struct. Biol.* **2008**, 161, 120.

[23] P.-Y. Chen, A. Y.-M. Lin, J. McKittrick, M. A. Meyers, *Acta Biomaterialia* **2008**, 4, 587.

[24] P. Compère, G. Goffinet, *Tissue and Cell* **1987**, 19, 859.

[25] J. McKittrick, P. Y. Chen, L. Tombolato, E. E. Novitskaya, M. W. Trim, G. A. Hirata, E. A. Olevsky, M. F. Horstemeyer, M. A. Meyers, *Materials Science and Engineering: C* **2010**, 30, 331.

[26] E. A. Zimmermann, R. O. Ritchie, *Advanced healthcare materials* **2015**, 4, 1286.

[27] B. Viswanath, R. Raghavan, U. Ramamurty, N. Ravishankar, *Scripta Materialia* **2007**, 57, 361.

[28] H. Fabritius, C. Sachs, D. Raabe, S. Nikolov, M. Friák, J. Neugebauer, in *Chitin*, Vol. 34 (Ed: N. S. Gupta), Springer Netherlands, **2011**, 35.

[29] L. K. Grunenfelder, S. Herrera, D. Kisailus, *Small* **2014**, 10, 3207.

[30] S. Weiner, W. Traub, H. D. Wagner, *Journal of Structural Biology* **1999**, 126, 241.

[31] W. Wagermaier, H. S. Gupta, A. Gourrier, M. Burghammer, P. Roschger, P. Fratzl, *Biointerphases* **2006**, 1, 1

[32] H. Lichtenegger, M. Muller, O. Paris, C. Riekel, P. Fratzl, *Journal of Applied Crystallography* **1999**, 32, 1127.

[33] E. A. Zimmermann, B. Gludovatz, E. Schaible, N. K. Dave, W. Yang, M. A. Meyers, R. O. Ritchie, *Nat. Commun.* **2013**, 4, 2634.

[34] M.-Y He, J. W. Hutchinson, *International Journal of Solids and Structures* 1989, 25, 1053.

[35] P. Fratzl, H. S. Gupta, F. D. Fischer, O. Kolednik, *Adv. Mater.* **2007**, 19, 2657.

[36] O. Kolednik, J. Predan, F. D. Fischer, P. Fratzl, *Adv. Funct. Mater*. **2011**, 21, 3634.

[37] R. Wang, H. S. Gupta, *Annual Review of Materials Research* **2011**, 41, 41.

[38] R. O. Ritchie, J. H. Kinney, J. J. Kruzic, R. K. Nalla, *Fatigue & Fracture of Engineering Materials & Structures* **2005**, 28, 345.

[39] P. D. Zavattieri, L. G. Hector, A. F. Bower, *International Journal of Fracture* **2007**, 145, 167.

[40] F. A. Cordisco, P. D. Zavattieri, L. G. Hector Jr, B. E. Carlson, *International Journal of Solids and Structures* **2016**, 83, 45.

[41] L. K. Grunenfelder, N. Suksangpanya, C. Salinas, G. Milliron, N. Yaraghi, S. Herrera, K. Evans-Lutterodt, S. R. Nutt, P. Zavattieri, D. Kisailus, *Acta Biomaterialia* **2014**, 10, 3997.

[42] L. K. Grunenfelder, E. E. de Obaldia, Q. Wang, D. Li, B. Weden, C. Salinas, R. Wuhrer, P. Zavattieri, D. Kisailus, *Adv. Funct. Mater.* **2014**, 24, 6093.

[43] T. H. Hsieh, A. J. Kinloch, K. Masania, A. C. Taylor, S. Sprenger, *Polymer* **2010**, 51, 6284

[44] B. Johnsen, A. Kinloch, R. Mohammed, A. Taylor, S. Sprenger, *Polymer* **2007**, 48, 530.

[45] P. Dittanet, R. A. Pearson, *Polymer* **2012**, 53, 1890.

[46] G. Dvorak, in *Micromechanics of composite materials*, Vol. 186, Springer Science & Business Media, **2012**.

[47] W. C. Oliver, G. M. Pharr, *Journal of Materials Research* **2004**, 19, 3.

[48] K. Moran, R. Wuhrer, *Microchimica Acta* **2006**, 155, 209.

[49] *Abaqus Standard: user's manual version 6.11.*, Simulia USA, 2011.

[50] A, Bower. *Applied mechanics of solids.* CRC press, 2009.

# Supplementary material

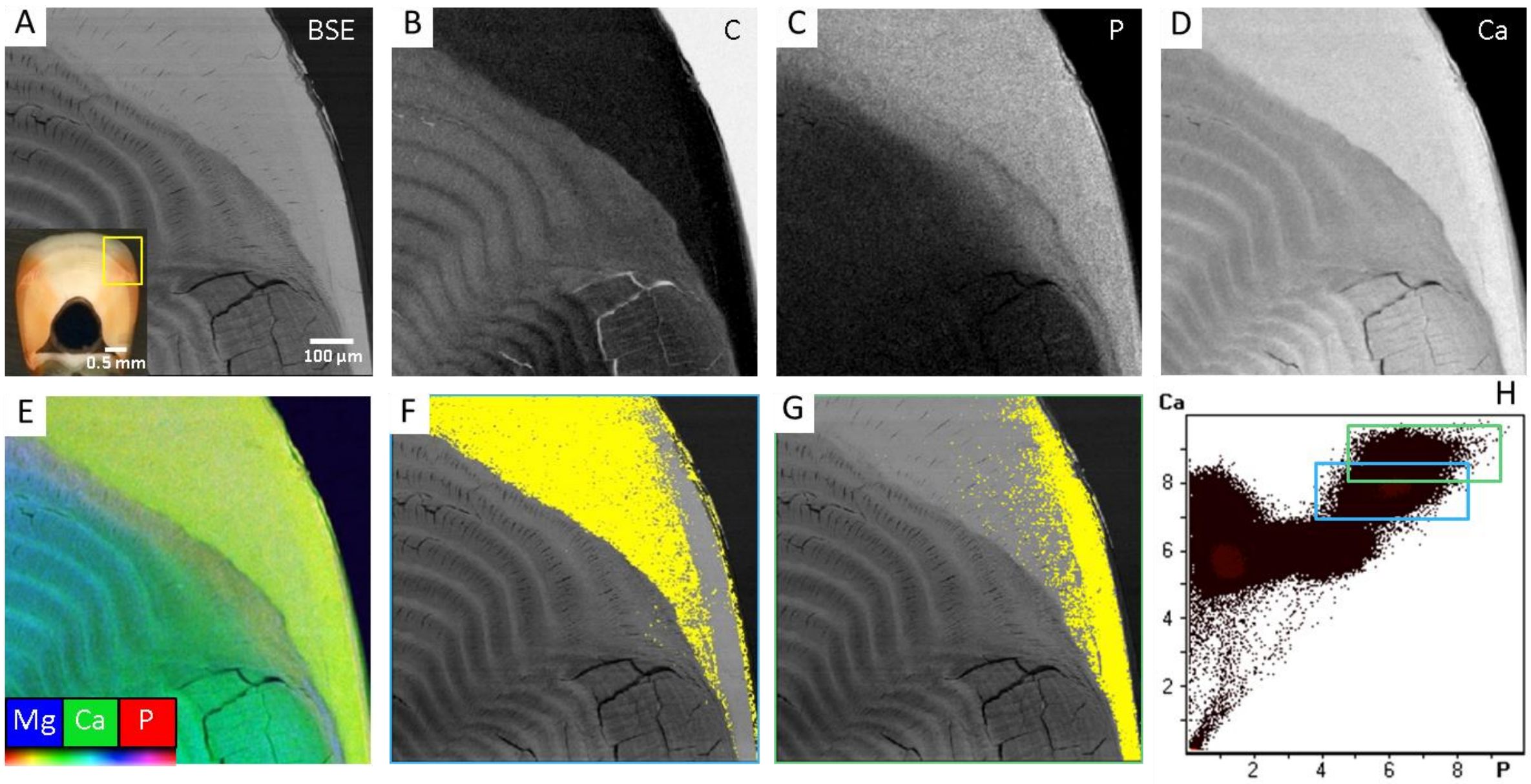

**Figure S1.** X-Ray mapping and elemental analysis of the impact region. (A) Backscattered electron micrograph of the impact region from a polished transverse section of the dactyl club. Inset showing polished transverse section with yellow boxed region indicating mapped region. (B-D) EDS maps showing local distribution of carbon, phosphorus, and calcium respectively. (E) Pseudo-colored map revealing local concentrations of magnesium (blue), calcium (green) and phosphorus (red). (F,G) Maps displaying regions of unique composition of calcium and phosphorus corresponding to the green and blue boxed areas respectively of the scatter diagram in (H). Highlighted region in (G) corresponds to nanoparticulated surface domain. (H) Scatter diagram of mapped area showing pixel frequency versus elemental concentration profiles for calcium and phosphorus plotted against one another.

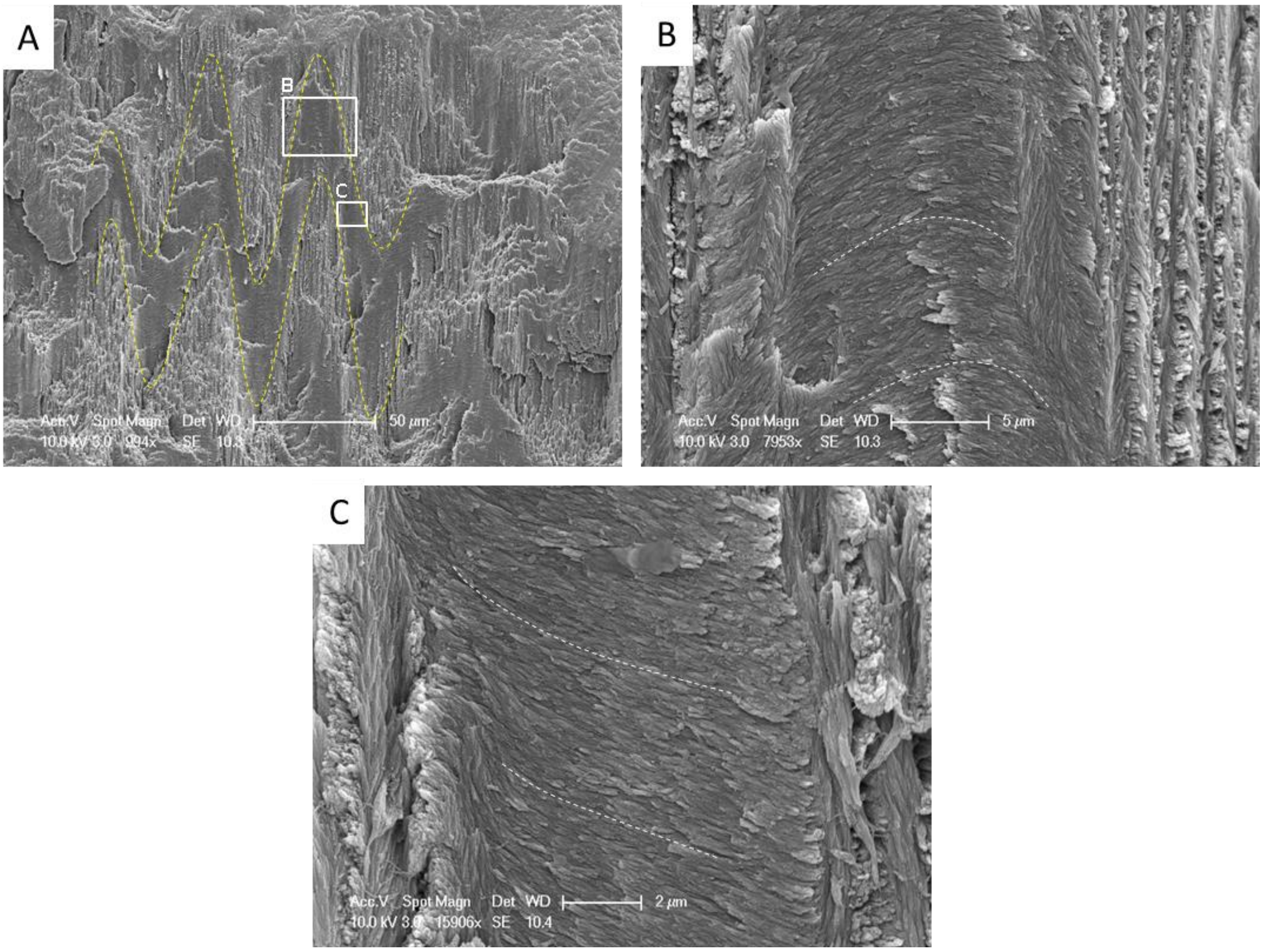

**Figure S2.** SEM analysis of a fractured sagittal section of the impact region. (A) Low magnification micrograph of the impact region. Dashed yellow lines highlight the herringbone structure and denote the transition between in-plane and out-of-plane fiber orientation. (B,C) High magnification micrographs from areas denoted in (A) showing in-plane fibers located at the peak of a "sinusoid" (B) and just above the sinusoid center-line (C). Dashed white lines showing in-plane fibers bent following the sinusoidal pattern of the herringbone structure.

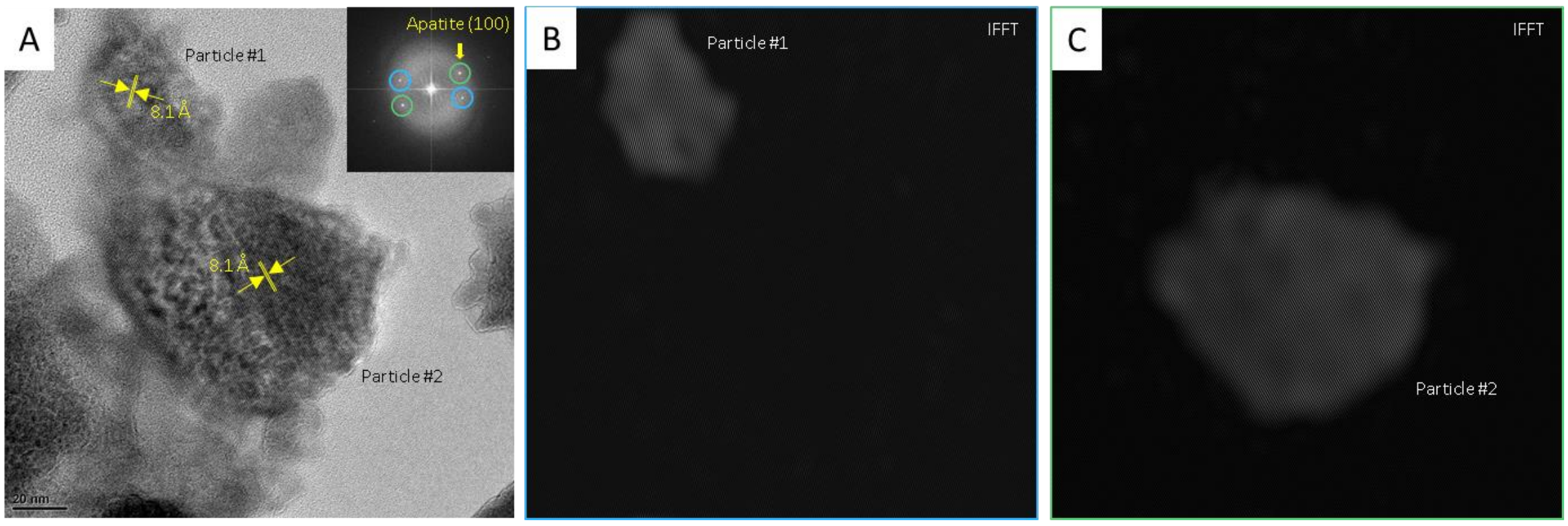

**Figure S3.** High-resolution transmission electron microscopy (HRTEM) and inverse fast Fourier-transform (IFFT) analysis of nanoparticles within impact surface. (A) HRTEM micrograph of two isolated nanoparticles within the impact surface. Inset: FFT highlighting two sets of misaligned apatite (100) lattice planes corresponding to particles #1 and #2. (B) Inverse FFT of masked (100) reflections (circled in blue) from the inset in (A) highlighting particle #1. (C) Inverse FFT of masked (100) reflections (circled in green) from the inset in (A) highlighting

particle #2. Inverse FFT analysis shows that the lattice planes extend through the entire particles, providing evidence of their single crystalline nature.

**Uniaxial strain analysis**

In the studied case the material properties for the finite element analysis are $E_t$ = 153.75 GPa, $E_p$ = 9.7 GPa, $\nu_p$ = 0.672, $\nu_t$ = 0.344, $G_{tp}$ = 5.79 GPa. $\rho$ = 1440 kg/m$^3$.[41] Where the subindex t refers to properties aligned with the axis of symmetry of the material and the subindex p refers to directions perpendicular to the axis of symmetry. The number of layers considered was 20, $x/\lambda \in [-2, 2]$, $y/\lambda \in [-2, 2]$, and $z/A \in [-20, 20]$. For the mesh, we used 81×81×81 trilinear elements (C3D8).[49] The body was subjected to uniaxial strain conditions in the $z$ direction. This analysis is used to compare the distribution of stresses in the microstructures for a given applied strain.

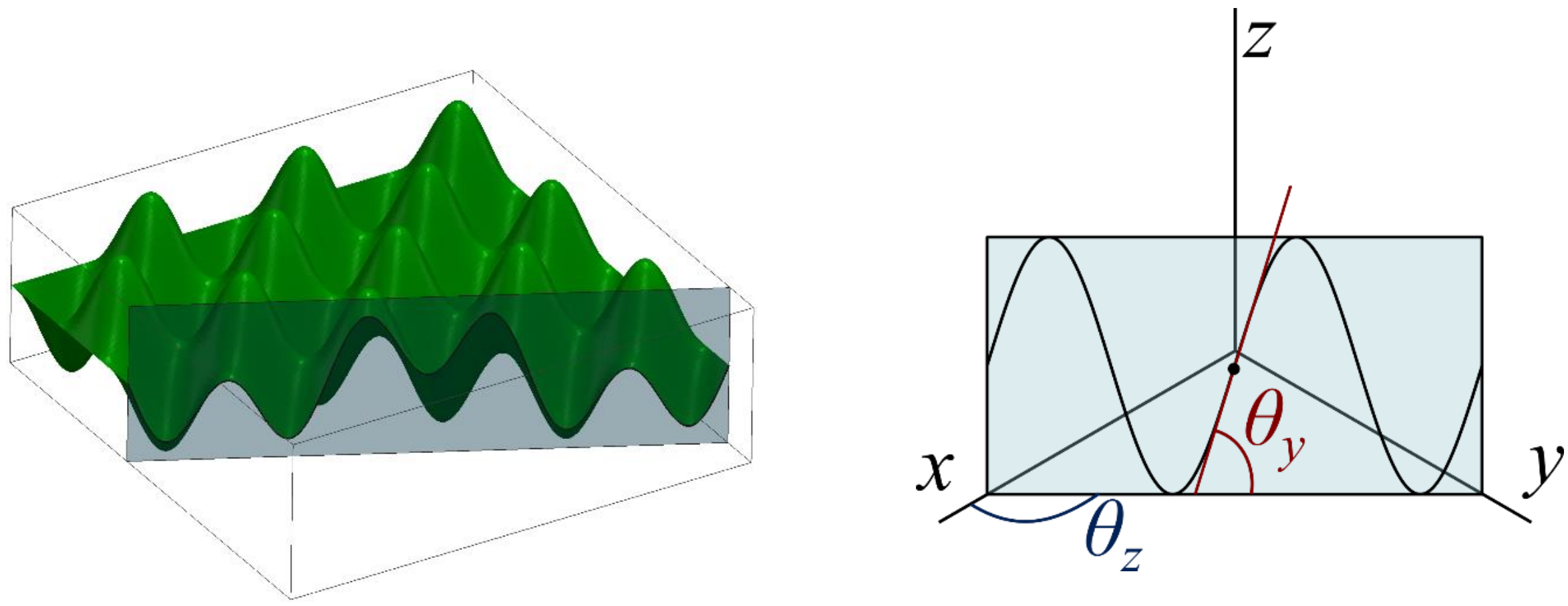

**Figure S4.** Fiber orientation generation over the eggshell surface.

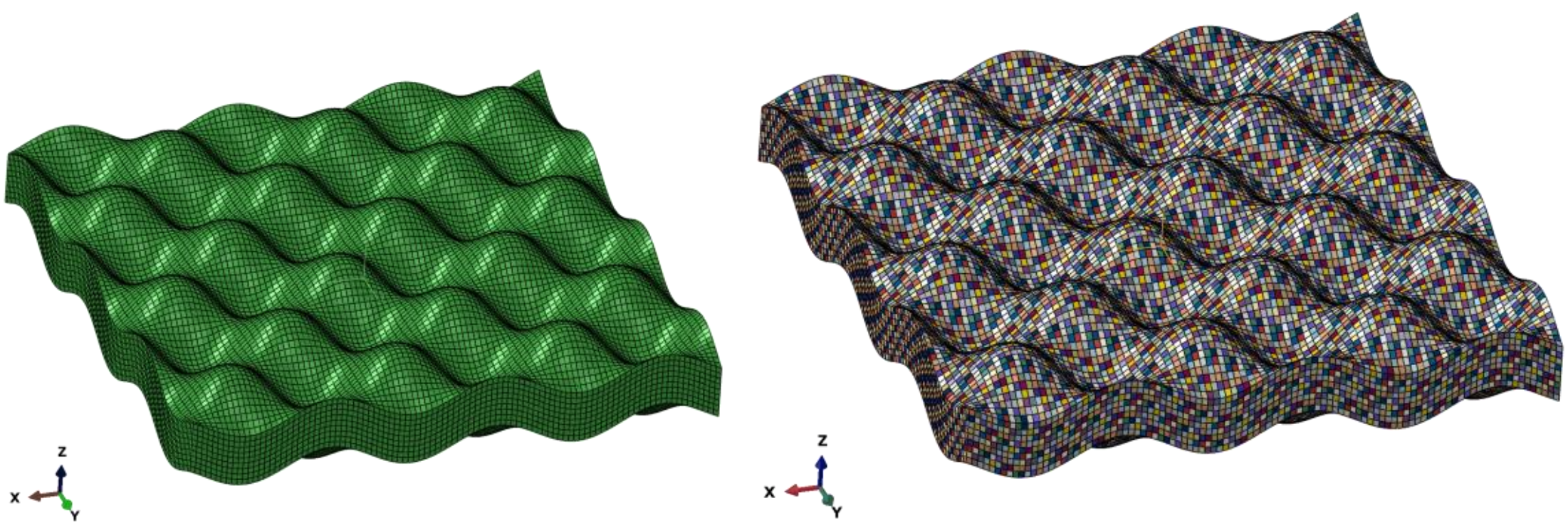

**Figure S5.** Mesh generated for a layer of the herringbone structure. The colored mesh depicts how the material properties change from element to element.

**3D Printing**

In both cases, the cylinders in Figure 5F are 60 mm in diameter and 37 mm in height. The fiber diameter is 0.6 mm, the horizontal spacing is 0.65 mm and the pitch angle is 18º. The vertical spacing for the Bouligand sample is 0.6 mm and the volume fraction 0.242. For the herringbone sample, the amplitude is 3 mm and the wavelength 2 mm. However, vertical spacing is slightly larger than the one in the Bouligand sample (i.e., 0.9 mm) to keep the same volume fraction, namely 0.267. For both materials, the fibers are printed with a hard polymer (VeroWhite) and the matrix with a soft elastomer (TangoBlack Plus).

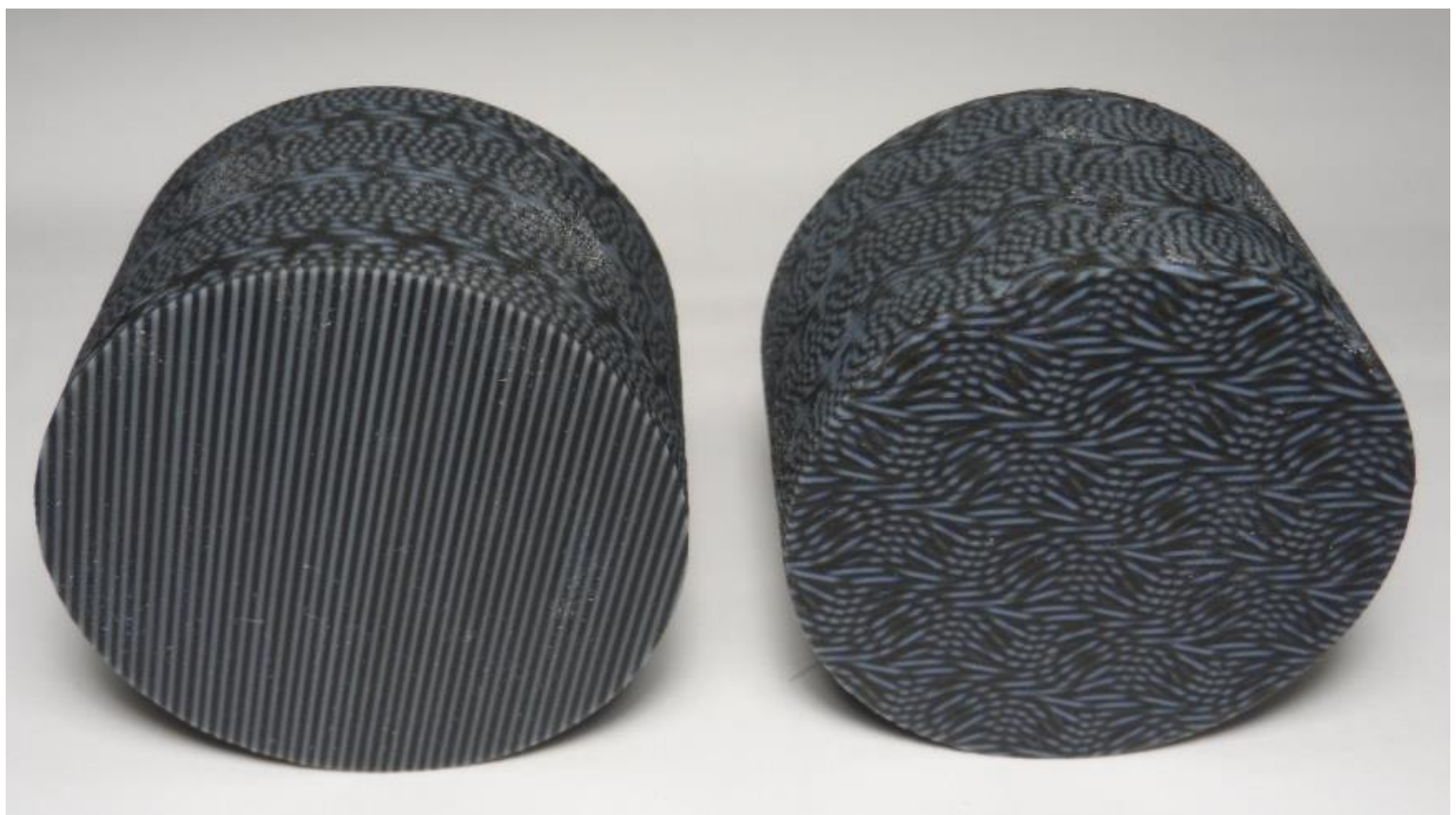

**Figure S6.** Overview image of 3D printed cylindrical Bouligand (left) and herringbone (right) samples.

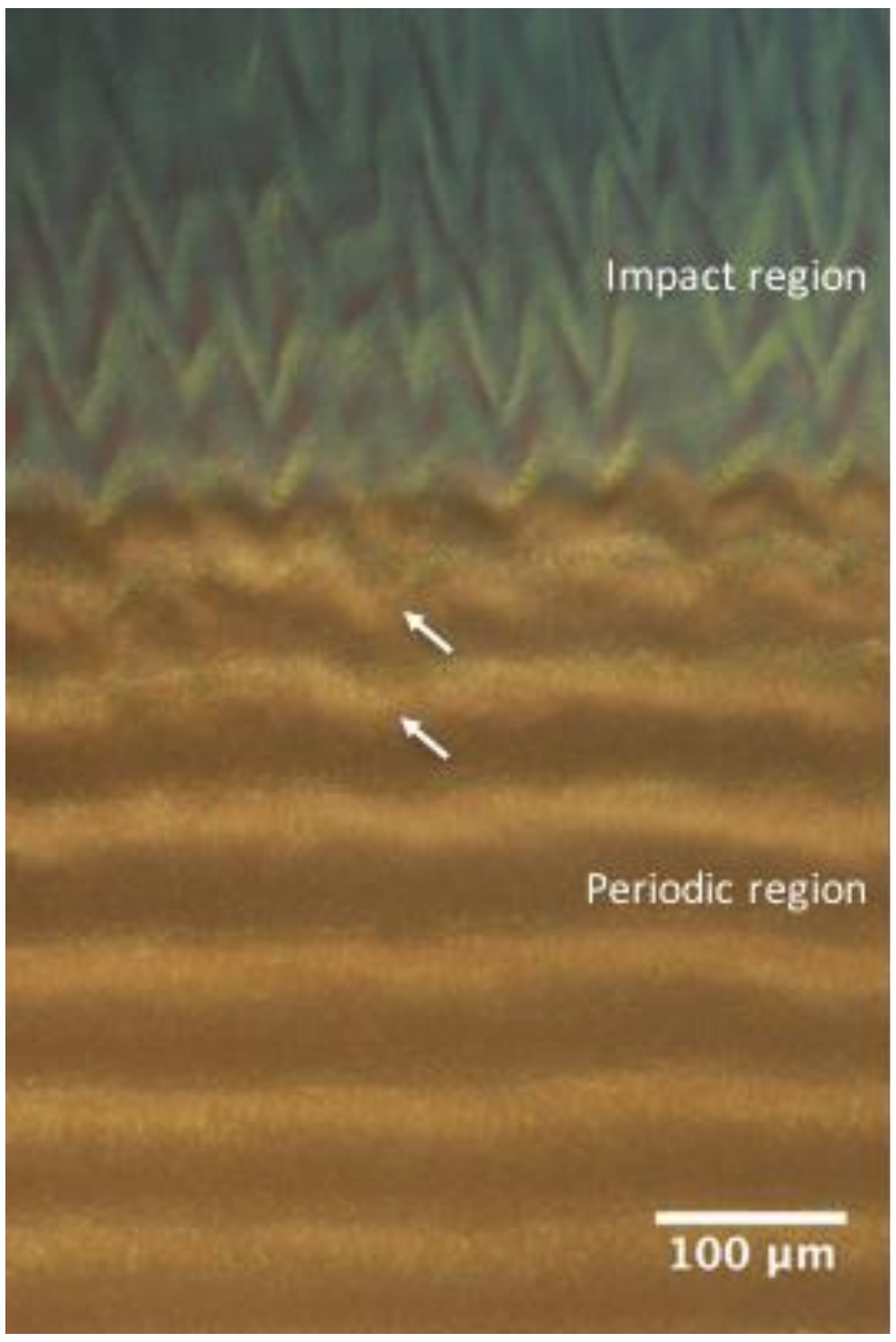

**Figure S7.** Differential interference contrast image of impact region-periodic region interface showing a gradual transition from flat Bouligand layers within the periodic region to a well defined herringbone structure within the impact region.

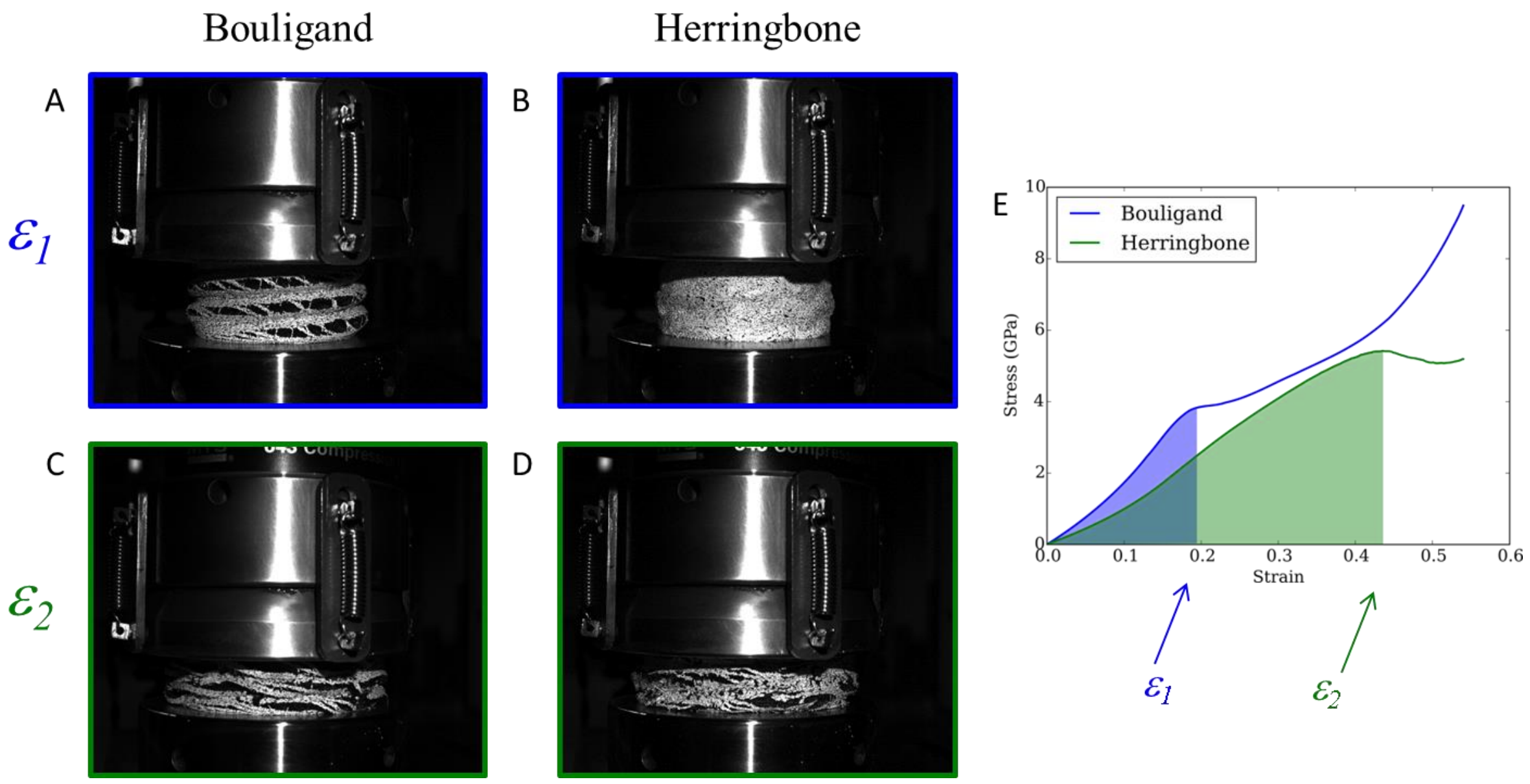

**Figure S8.** Comparison of compression testing on 3D printed biomimics at two different strains. Optical micrographs of Bouligand (A) and herringbone (B) structures corresponding to strain of $\varepsilon_1 \approx 0.20$ as shown in (E). Note that the Bouligand structure has undergone significant damage in comparison to the herringbone structure. Optical micrographs of the Bouligand (C) and herringbone (D) structures corresponding to strain of $\varepsilon_2 \approx 0.44$ as shown in (E). Both structures have undergone significant damage. (E) Stress-strain diagram showing results of the compression testing. Excessive damage within the Bouligand and herringbone structures is initiated at $\varepsilon_1$ and $\varepsilon_2$, respectively. Subsequent behavior likely corresponds to hardening of the structures due to compaction of the fibers.